\documentclass[twocolumn,prl,showpacs,floatfix,amsfonts,superscriptaddress, bibnotes]{revtex4} 
 
\usepackage[]{graphicx} 
\usepackage{amsbsy} 
\usepackage{float}
 
\begin{document} 
\title{Ultrafast surface carrier dynamics in the topological insulator Bi$_2$Te$_3$}

\author{M.~Hajlaoui} 
\affiliation{Laboratoire de Physique des Solides, CNRS-UMR 8502, Universit\'{e} Paris-Sud, F-91405 Orsay, France}
\author{E.~Papalazarou} 
\affiliation{Laboratoire de Physique des Solides, CNRS-UMR 8502, Universit\'{e} Paris-Sud, F-91405 Orsay, France} 
\author{J.~Mauchain} 
\affiliation{Laboratoire de Physique des Solides, CNRS-UMR 8502, Universit\'{e} Paris-Sud, F-91405 Orsay, France} 
\author{G.~Lantz} 
\affiliation{Laboratoire de Physique des Solides, CNRS-UMR 8502, Universit\'{e} Paris-Sud, F-91405 Orsay, France} 
\author{N.~Moisan} 
\affiliation{Laboratoire d'Optique Appliqu\'{e}e, ENSTA, CNRS, Ecole Polytechnique, 91761 Palaiseau, France}
\author{D.~Boschetto} 
\affiliation{Laboratoire d'Optique Appliqu\'{e}e, ENSTA, CNRS, Ecole Polytechnique, 91761 Palaiseau, France}
\author{Z.~Jiang} 
\affiliation{School of Physics, Georgia Institute of Technology, Atlanta, Georgia 30332, USA} 
\author{I.~Miotkowski} 
\affiliation{Department of Physics, Purdue University, West Lafayette, IN 47907, USA}
\author{Y.P.~Chen} 
\affiliation{Department of Physics, Purdue University, West Lafayette, IN 47907, USA}
\author{A.~Taleb-Ibrahimi}
\affiliation{Synchrotron SOLEIL, Saint-Aubin BP 48, F-91192 Gif-sur-Yvette, France}
\author{L.~Perfetti} 
\affiliation{Laboratoire des Solides Irradi\'{e}s, Ecole Polytechnique-CEA/DSM-CNRS UMR 7642, F-91128 Palaiseau, France} 
\author{M.~Marsi}
\affiliation{Laboratoire de Physique des Solides, CNRS-UMR 8502, Universit\'{e} Paris-Sud, F-91405 Orsay, France}

\date{\today} 
 
\begin{abstract} 

We discuss the ultrafast evolution of the surface electronic structure of the topological insulator Bi$_2$Te$_3$ following a femtosecond laser excitation.
Using time and angle resolved photoelectron spectroscopy, we provide a direct real-time visualisation of the transient carrier population of both the surface states and the bulk conduction band. We find that the thermalization of the surface states is initially determined by interband scattering from the bulk conduction band, lasting for about 0.5 ps; subsequently, few ps are necessary for the Dirac cone non-equilibrium electrons to recover a Fermi-Dirac distribution, while their relaxation extends over more than 10 ps.
The surface sensitivity of our measurements makes it possible to estimate the range of the bulk-surface interband scattering channel, indicating that the process is effective over a distance of 5 nm or less. This establishes a correlation between the nanoscale thickness of the bulk charge reservoir and the evolution of the ultrafast carrier dynamics in the surface Dirac cone.  

%
%
%
%
\end{abstract} 


\maketitle


Topological insulators are new states of quantum matter that are attracting a lot of interest due to their unique transport properties~\cite{Fu2007, Qi2008}. A distinctive feature of these materials is the fact that they have conducting states at their edge or surface which are protected by time-reversal symmetry: as a consequence, they cannot be adiabatically connected to the bulk which is instead insulating~\cite{Hasan2010, Qi2011}. 
Examples of three-dimensional topological insulators that present these properties - with their metallic surface bands constituting a Dirac cone - are Bi$_2$Se$_3$, Sb$_2$Te$_3$ and Bi$_2$Te$_3$~\cite{Zhang2009}.  
These properties make them natural candidates for a new generation of electronic devices~\cite{Wang2012}, but for such a development to be realistic a description of the dynamical response of these materials and a detailed knowledge of their conduction electronic states are needed. Furthermore, since the Dirac cone 
is a surface feature, it is essential to determine its interband scattering range, i.e. the distance over which the surface bands effectively exchange carriers with the bulk. 

\begin{figure}[hb] 
\includegraphics[width=1\linewidth,clip=true]{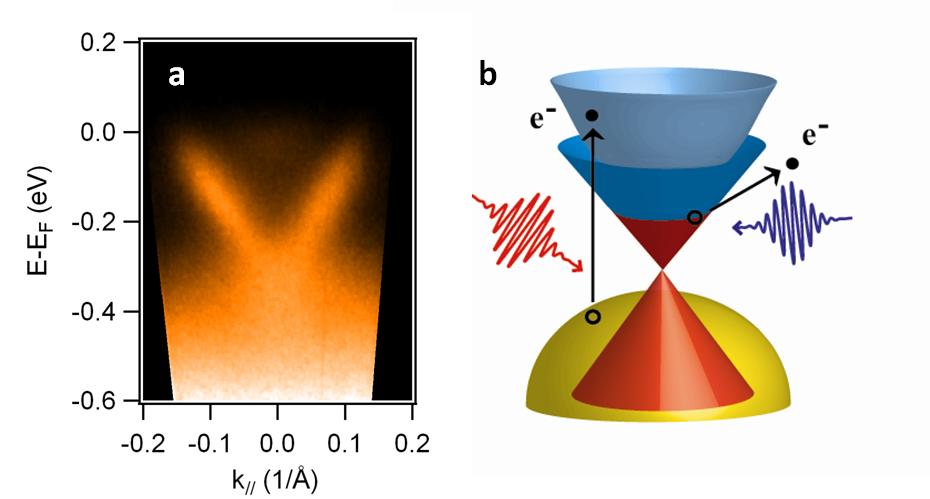} 
\caption{(a) ARPES image acquired along the $\Gamma K$ direction in s-polarization with the 6.3 eV laser source. (b) pictorial view of the experiment: the infrared laser pump excites electrons from the sample valence band (orange) into the empty conduction band (light blue). The subsequent carrier flow produces a transient variation in the charge population in the surface Dirac cone, both for empty (blue) and filled (red) electronic states, and is probed by the ultraviolet probe.} 
\label{arpes}
\end{figure}

\begin{figure*}[htb] 
\includegraphics[width=1\linewidth,clip=true]{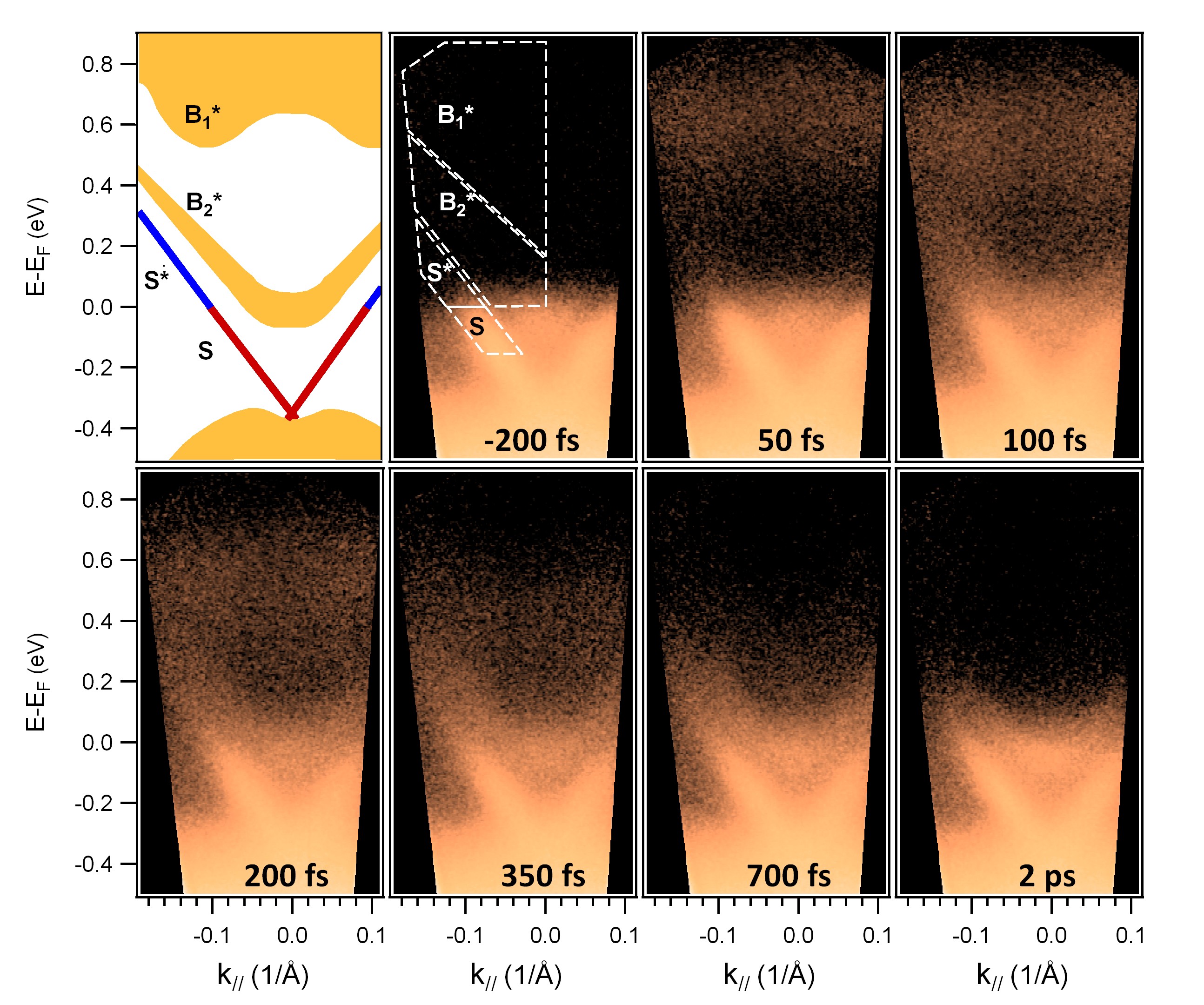} 
\caption{Time-resolved ARPES sequence obtained after photoexcitation from the pump pulses: the color intensity is presented in logarithmic scale to make the signal from the transient electronic states more evident. In the upper left corner, a schematic view of the bands involved in the process: the projections of the bulk bands $B_1^*$ and $B_2^*$, and the surface Dirac cone. In the first image (t=-200 fs) the integration regions $S$, $S^*$, $B_1^*$ and $B_2^*$ are defined} 
\label{arpes_vs_t}
\end{figure*}

In this letter, we show how pump-probe ultrafast Angle Resolved PhotoElectron Spectroscopy (ARPES) can be used to answer these questions. ARPES is playing a key role in the study of topological insulators, thanks to its ability to map electronic states in k-space and to its surface sensitivity~\cite{Hsieh2008,Chen2009,Xia2009, Hsieh2009}. 
The same unique advantages, coupled with the femtosecond resolution achievable with laser sources in pump-probe configuration, make time-resolved ARPES the technique of choice to explore the transient electronic structure and the evolution of out-of-equilibrium topological insulators. In the case of Bi$_2$Te$_3$, we found that the bulk conduction band acts as a charge reservoir 
that reaches a quasi-equilibrium condition with the surface bands in about 0.5 ps and allows us to estimate the ballistic range for electrons travelling within the surface in the edge bands (180 nm). The hot electrons in the Dirac cone evolve towards a Fermi-Dirac distribution over few ps, while their subsequent relaxation process, lasting for more than 10 ps, is considerably longer than the main bulk recombination processes.  
Furthermore, the comparison of our time resolved ARPES results with ultrafast optical reflectivity measurements on the same material indicates that in Bi$_2$Te$_3$ the interband scattering of carriers between bulk and surface states is effective over a distance perpendicular to the surface of 5 nm. This can be a relevant piece of information in possible future device applications based on thin films of this material, since it means that the film thickness can affect the ultrafast response of the Dirac cone, representing a potential tuning parameter to control carrier dynamics to and from the surface topological bands. 

Single crystals of Bi$_2$Te$_3$ (n-type) were carefully oriented and cleaved under ultrahigh vacuum conditions in the recently commissioned FemtoARPES experimental setup~\cite{Papalazarou2012,Faure2012}, where pump-probe photoemission measurements can be performed with a time resolution of 65 fs and an energy resolution of 60 meV. Photoelectrons were excited with 6.32 eV photons, obtained as the 4th harmonic of the 1.58 eV laser which is used to pump the transient states in pump-probe configuration. The use of a 6.32 eV probe gives access to a limited portion of $k$-space, but sufficient in this case to observe the Dirac cone produced by the surface bands. In Fig.~\ref{arpes}(a) we present a conventional (i.e. without the pump laser) ARPES image along the $\Gamma K$ direction. The results are consistent with previous ARPES studies at higher photon energy on Bi$_2$Te$_3$~\cite{Chen2009}: the surface Dirac cone is clearly visible. The probing depth of low energy ARPES can be estimated to about 2-3 nm ~\cite{Rodolakis2009}. This optimizes the simultaneous detection of the surface dynamics (in particular the Dirac cone) and of the bulk dynamics which is probed by ultrafast optical reflectivity measurements on Bi$_2$Te$_3$ ~\cite{Wu2008,Kamaraju2010}. The sibling compound Bi$_2$Se$_3$ was also studied with ultrafast optical spectroscopy~\cite{Qi2010,Hsieh2011}, and selective sensitivity to the surface was  achieved using second harmonic reflections~\cite{Hsieh2011}. With fs pump-probe ARPES, one has direct access to the transient electron population and its density of states: in this work we focused on these aspects, paying particular attention to the sub-ps thermalization regime. 

In this study we present results obtained on the $\Gamma K$ direction, where no hybridization is present between the surface Dirac cone and the bulk conduction band states in the proximity of the 
Fermi energy $E_{F}$ ~\cite{Chen2009,Xia2009,Zhang2009}: this made it possible to unambiguously discriminate the photoemission yield between the empty surface (blue in Fig.~\ref{arpes}(b)) and bulk (light blue in Fig.~\ref{arpes}(b)) bands over an extended wavevector-energy window. We explored also other directions in $k$-space (namely $\Gamma M$), and no relevant difference was found in the observed carrier dynamics, indicating that the hexagonal deformation of the Dirac cone present in Bi$_2$Te$_3$ does not play a critical role in the ultrafast response of the system. All the data presented here were taken at room temperature.  


In Fig.~\ref{arpes_vs_t} we present the evolution of the ARPES yield at different pump-probe time delays. The schematic band structure of Bi$_2$Te$_3$ shown in the upper left corner of Fig.~\ref{arpes_vs_t} is consistent with recent theoretical  calculations~\cite{Zhang2009,Yazyev2010}, and can be used as a guide to the eye to interpret our data.
For sake of simplicity, during our study we followed the 
evolution of the electron population of the regions in $k$-space shown for the t=-200 fs time delay in Fig.~\ref{arpes_vs_t}: one for each projection of the bulk conduction bands $B_1^*$ and $B_2^*$, 
one for the occupied surface band $S$ and one for the unoccupied surface band $S^*$.  In particular, the separation between the conduction band $B_2^*$ and the surface band $S*$ is clear from the experimental data (for instance at t=700 fs), so that our choice of the integration regions makes it possible to unambiguously follow the evolution of the different bands.  
This approach allowed us to capture the essential features of the ultrafast carrier dynamics in the nanometer region close to the surface.  

\begin{figure}[htb] 
\includegraphics[width=1\linewidth,clip=true]{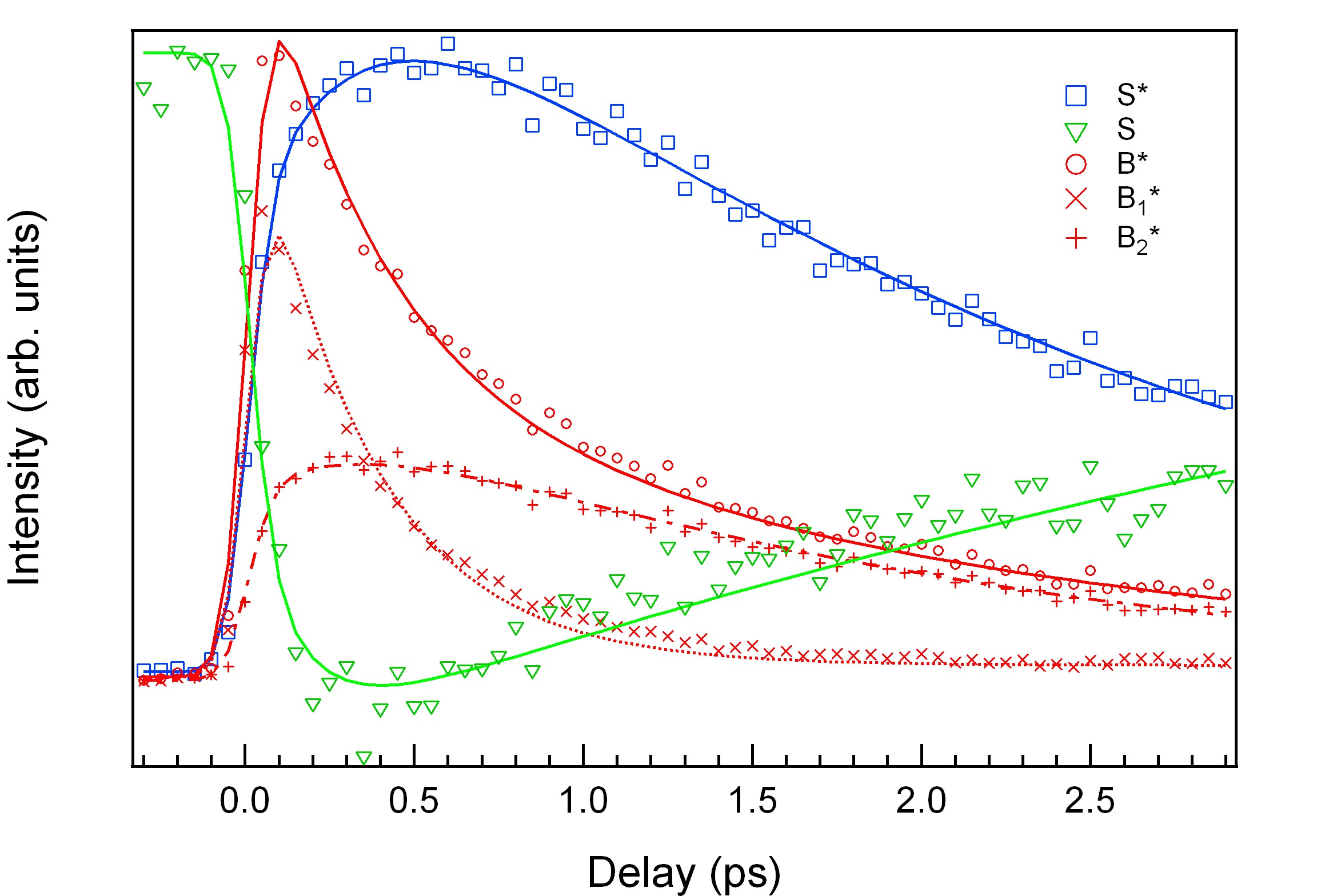} 
\caption{Ultrafast evolution of the populations of $S$, $S^*$, ${B^*_{1}}$ and ${B^*_{2}}$. The fitting curves (solid lines) were obtained as explained in the text.} 
\label{populations}
\end{figure}

Following the pump excitation (50 fs, 1.58 eV, 0.15 $mJ/cm^{2}$), a transient excited electron population is observed in the conduction band $B_1^*$ around 0.6 eV above $E_{F}$ (t=100 fs). This transient population relaxes quickly (in about 300 fs), triggering a cascade of recombination processes, and acting as a source of electrons for the bulk conduction band $B_2^*$ and for the surface bands. This is well visible also in the movie provided with the supplementary material. The images in Fig.~\ref{arpes_vs_t} provide a direct visualisation of the electron dynamics, but in order to get more quantitive information, we also provide in Fig.~\ref{populations} the time evolution of the populations of $S$, $S^*$, $B_1^*$ and $B_2^*$, as well as of the overall excited bulk population $B^*(t)=B_1^*(t) + B_2^*(t)$.

\begin{figure}[htb] 
\includegraphics[width=1\linewidth,clip=true]{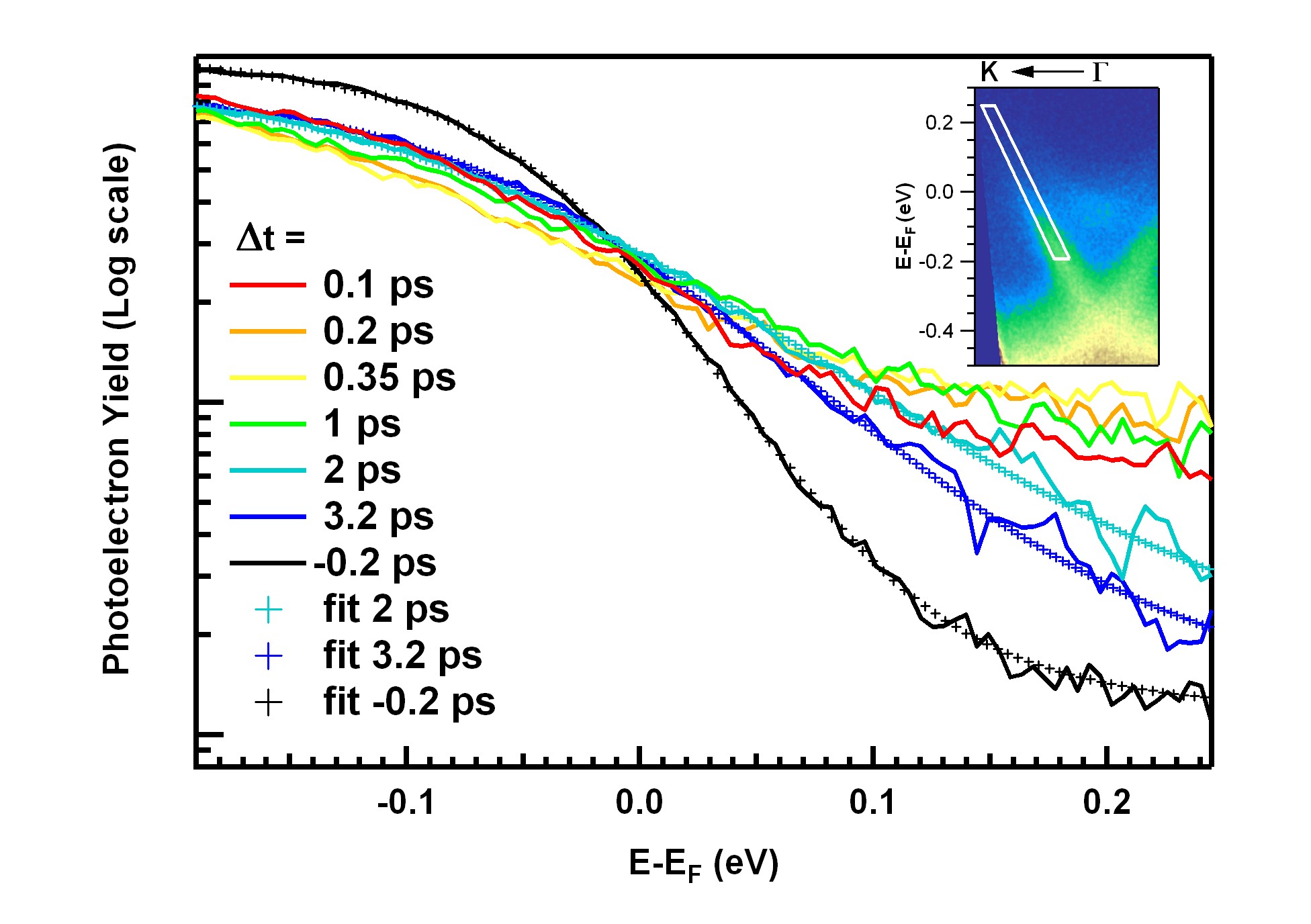} 
\caption{Energy Distribution Curves along the surface state $S+S^*$ at different pump-probe delays. For the spectra at -0.2, 2 and 3 ps we also present the corresponding fits with a Fermi-Dirac distribution. The photoemission yield is presented in logarithmic scale. In the inset, the 
corresponding momentum integration area is shown.} 
\label{EDC}
\end{figure}

The ultrafast carrier dynamics between the surface ($S$, $S^*$) and the bulk  ($B^*$) can be described using a standard rate equation model for semiconductors~\cite{Marsi1998,Halas1989} to fit the population curves $S^*(t)$, $S(t)$ and to identify the main charge flow processes to and from the surface bands: for instance, femtosecond carrier dynamics could be successfully studied at the surface of III-V semiconductors, as clearly demonstrated for the surface resonance of InP(100)~\cite{Toben2005}. Nevertheless, in the case of Bi$_2$Te$_3$ one is not confronted with weakly dispersing surface resonances, normally found in standard semiconductors: on the contrary, like for all topological insulators the surface bands are metallic, strongly dispersing and cross the whole bulk energy gap. This makes it necessary to adopt an alternative approach to study their relaxation, rather than introducing in the rate equations recombination terms proportional to the product $S(t)S^*(t)$) as normally done for the description of semiconductor surface resonances. As we will see later (Fig.~\ref{EDC}), 
the detailed study of the evolution of the photoelectron Energy Distribution Curves (EDC') of the Dirac cone can represent such an alternative approach.  


Based on the assumption that $B_1^*$ acts as a source of excess electrons that adds up to direct excitation for $B_2^*$ and $S^*$ with relative weights ${\alpha}$ and ${\beta}$, respectively,   
the main carrier interband scattering processes can be described by the following equations: 

\begin{eqnarray}
\frac{dB^*_1}{dt}=G_{1}(t) -\frac{B^*_1(t)}{\tau_1}\\
\frac{dB^*_2}{dt}=G_{2}(t) +{\alpha}\frac{B^*_1(t)}{\tau_{1}}-\frac{B^*_2(t)}{\tau_2}\\
\frac{dS^*}{dt}=G_{S^*}(t) +{\beta}\frac{B^*_1(t)}{\tau_{1}}-\frac{S^*(t)}{\tau_{D1}}\\
\frac{dS}{dt}=
-G_{S}(t)-\frac{B_{tot}-B(t)}{\tau_{h}}+\frac{S_{tot}-S(t)}{\tau_{D2}}
\end{eqnarray}

the flow of charge to $S^*(t)$ is described by direct excitation from the laser pulse $G_{S*}$, by a flow of electrons from the ${B^*_1(t)}$ projection of the bulk conduction band and by the recombination processes for excess electrons described by $\tau_{D1}$. Similarly, the flow of charge to ${S(t)}$ is described by direct excitation from the laser pulse $G_{S}$ (gaussian with the same FWHM), by a flow of holes from the bulk valence band $(B_{tot}-B(t))$ with characteristic time ${\tau_{h}}$, and by the recombination processes for excess holes $(S_{tot}-S(t))$ described by $\tau_{D2}$. ${B^*_2(t)}$ can be also described by direct excitation, by interband scattering from ${B^*_1}$, and by a decay time $\tau_{2}$.
We used the solutions of these coupled differential equations as fitting functions for our population curves ${B^*_1(t)}$, ${B^*_2(t)}$, $S^*(t)$ and $S(t)$, obtaining as a result the curves shown as solid lines in Fig.~\ref{populations} for the following values of the fitting parameters: FWHM of 76 fs for $G_{1}$, $G_{2}$, $G_{S*}$, and $G_{S}$; ${\alpha}=0.18$; ${\beta}=0.12$; ${\tau_{1}}=0.34 ps$; ${\tau_{2}}=1.8 ps$; $\tau_{D1}=2.1 ps$; ${\tau_{h}}=0.14 ps$; $\tau_{D2}=5.7 ps$. 

There are obviously many different phenomena that contribute to the different characteristic times presented here (electron-phonon scattering, carrier trapping at defects, etc.), both for surface and for bulk bands. We adopted a simplified description (surface vs bulk) because our main interest here is to determine the scattering time between the Dirac cone and the bulk bands: as we have seen, the fast decay of ${B^*_1}$ is strictly related to the time needed for both $S$ and $S^*$ to reach the maximum population change (Fig.~\ref{populations}). 
$S(t)$ and $S^*(t)$ present nevertheless a slightly asymmetric behavior while reaching a thermalization regime, and also during the relaxation of the excess carriers, as indicated by the different values of $\tau_{D1}$ and $\tau_{D2}$. Electron-hole recombination between $S$ and $S^*$ gives of course a symmetric decay term, consequently the asymmetry must be due to the presence of decay terms which are markedly different between electrons and holes (like scattering to defects). The asymmetric flow of electrons and holes on the ps and longer time scale from surface states is frequently found in many other semiconducting materials, and can be for instance simply related to the inequivalent position of the surface bands with respect to the valence and conduction bands~\cite{Marsi1997}: we shall not discuss this here, but we found that longer scans show that after about 15 ps both $S^*(t)$ and $S(t)$ are back to their values before excitation. Furthermore, the very similar values of $\tau_{D1}$ and $\tau_{2}$ indicate that $S^*(t)$ and $B_2(t)$ present a parallel evolution: even though we didn't explicitely insert a scattering term between them in the rate equation, this similarity indicates that an effective carrier exchange takes place between the two bands. After completion of our work, we became aware of a time-resolved ARPES experiment on p-type Bi$_2$Se$_3$~\cite{Sobota2012}, where the authors also found a transient surface state population persisting for more than 10 ps, due to the reservoir role played by the bottom of the conduction band. The fact that different materials with different doping levels show this effect suggests that a slow relaxation of surface hot electrons may be a general property of topological insulators. 

Following the temporal evolution of the populations gives a good description of the charge flow, but more precise information can be obtained on the electronic structure of the photoexcited surface states by looking at the details of the photoelectron EDC's. In Fig.~\ref{EDC} we present the EDC's extracted from the surface bands (the momentum integration regions are visible in the inset) at some representative time delays: these curves can give a direct description of the Dirac cone during its thermalization process. One can see that for the longer delays, namely 2, 3 and -0.2 ps after the pump-pulse, the EDC's can be well fitted with a Fermi-Dirac distribution (the spectra at 2 and 3 ps give an electronic temperature of 792 K and 619 K, respectively). For shorter time delays, i.e. 0.1, 0.2 and 0.4 ps, the EDC's cannot be fitted with a Fermi-Dirac distribution, due to a strong presence of electrons at high energy, indicating that the system is not thermalized yet: at these delays the population of $S^*$ is still increasing due to the flow of charge from the bulk bands. This confirms that the dynamics of the surface Dirac cone is somehow delayed with respect to the bulk conduction band dynamics. Furthermore, if one multiplies the thermalization time (0.5 ps) by the Fermi velocity estimated from the ARPES data (0.36 m/$\mu$s), one finds that the scattering distance for electrons travelling parallel to the surface is of the order 180 nm: this sets a length scale for estimating the ballistic range for transport in the topological bands.

We should now discuss some implications of these results, also comparing our pump-probe ARPES data with ultrafast reflectivity in the optical range on Bi$_2$Te$_3$. The two techniques are very complementary in terms of surface sensitivity: while the probing depth of low energy ARPES is about 2-3 nm~\cite{Rodolakis2009}, optical reflectivity is sensitive to more than 10 nm~\cite{Wu2008}. One should also keep in mind that while our $B^*$ is a signal given only by the conduction band population, optical reflectivity probes the joint density of states from both the valence and the conduction band. Ultrafast reflectivity measurements performed by ourselves on the same single crystals studied with photoemission as well as by other authors on thin films of Bi$_2$Te$_3$~\cite{Wu2008} at comparable fluences make it possible to evaluate the carrier density at different depth values $z$ from the surface (see Fig. 4 in~\cite{Wu2008}). A fast decay time can be identified in the carrier density evaluated from reflectivity data~\cite{Wu2008}, which has been attributed to electron-phonon scattering (0.5 ps) and to carrier diffusion (more than 1 ps). It should be noted that even from reflectivity, the sub-ps carrier decay is estimated to be more pronounced at z=0 than at z=10 nm (Fig. 4 in~\cite{Wu2008}). The comparison with our data shows that in $B^*(t)$ the sub-ps decay is much more pronounced due to the charge flow to $S^*$, with ${\tau_{e}}=0.35$ ps - mainly because ARPES is more surface sensitive. This makes it possible to obtain a quantitative estimate of the distance from the surface over which scattering for excess electrons from the conduction band to the surface bands is effective. If one assumes an exponential attenuation of the photoemission yield $I(z)$,  $I(z)=I_{0}exp(-z/{\mu})$, one finds that for a surface signal to be comparable to a bulk one, it must be originated from a slab of thickness $\Delta z$ comparable to $\mu$ (2 nm); furthermore, it must certainly be less than the sensitivity of optical reflectivity (10 nm).  We can thus conclude that in Bi$_2$Te$_3$ the scattering of hot electrons from the conduction band to the Dirac cone is confined to a surface layer with a thickness of less than 5 nm. 
This means that altering the thickness of the subsurface region on the nanometer scale will affect the transient carrier dynamics in the Dirac cone, providing an important element and a potential tailoring parameter for possible technological applications of Bi$_2$Te$_3$ and of other topological insulators. In particular, the knowledge of the carrier scattering between the surface bands and the bulk is critically important to extend to topological insulators some technological possibilities demonstrated for truly bidimensional systems like graphene~\cite{Li2012}, related to ultrafast population inversion in the Dirac cone.  


In conclusion, we studied the transient electronic dynamics at the surface and subsurface region of the topological insulator Bi$_2$Te$_3$. Using time resolved ARPES, we provided a direct visualisation of the excess carrier population and of its evolution. We found that the ultrafast dynamics of the carriers in the surface Dirac cone is delayed with respect to the bulk, because the bulk acts as a reservoir that keeps providing a relevant charge flow for more than 0.5 ps, and because the subsequent relaxation phase (more than 10 ps), governed by charge diffusion and electron-phonon coupling, is less efficient than the typical bulk recombination. The surface sensitivity of our measurements make it possible to provide a quantitative estimate of the distance over which the bulk-surface interband carrier scattering is effective (about 5 nm). The results presented here for Bi$_2$Te$_3$ appear of general interest also for other three-dimensional topological insulators, in particular for possible nanoscale technological applications of these materials. 
 
Materials synthesis at PU is supported by the DARPA MESO program. 
Z.J. acknowledges support from the DOE (DE-FG02-07ER46451). 
The FemtoARPES project was funded by the RTRA Triangle de la Physique, the Ecole Polytechnique and the ANR (grant Nr. ANR-08-CEXCEC8-011-01). 



\end{document}